\documentclass[12pt,preprint]{aastex62}

\shortauthors{Bower et al.}
\shorttitle{Sgr A* THz Spectrum}
\begin{document}

\newcommand\degd{\ifmmode^{\circ}\!\!\!.\,\else$^{\circ}\!\!\!.\,$\fi}
\newcommand{\etal}{{\it et al.\ }}
\newcommand{\uv}{(u,v)}
\newcommand{\rdm}{{\rm\ rad\ m^{-2}}}
\newcommand{\msun}{{\rm\ M_{\sun}}}
\newcommand{\msuny}{{\rm\ M_{\sun}\ yr^{-1}}}
\newcommand{\mylesssim}{\stackrel{\scriptstyle <}{\scriptstyle \sim}}
\newcommand{\lsim}{\stackrel{\scriptstyle <}{\scriptstyle \sim}}
\newcommand{\gsim}{\stackrel{\scriptstyle >}{\scriptstyle \sim}}
\newcommand{\sci}{Science}
\newcommand{\sgr}{PSR J1745-2900}
\newcommand{\sgra}{Sgr~A*}
\newcommand{\kms}{\ensuremath{{\rm km\,s}^{-1}}}
\newcommand{\kkms}{\ensuremath{{\rm K\,km\,s}^{-1}}}
\newcommand{\masy}{\ensuremath{{\rm mas\,yr}^{-1}}}
\newcommand{\frb}{FRB 121102}
\newcommand{\ergs}{\ensuremath{{\rm erg\,s}}^{-1}}

\def\kbar{{\mathchar'26\mkern-9mu k}}
\def\totd{{\mathrm{d}}}


\title{ALMA Observations of the Terahertz Spectrum of Sagittarius A*}
\author[0000-0003-4056-9982]{Geoffrey C.\ Bower}
\affiliation{Academia Sinica Institute of Astronomy and Astrophysics, 645 N. A'ohoku Place, Hilo, HI 96720, USA}
\email{gbower@asiaa.sinica.edu.tw}

\author[0000-0003-3903-0373]{Jason Dexter}
\affiliation{Max Planck Institute for Extraterrestrial Physics, Giessenbachstrasse 1, D-85748 Garching, Germany}
\affiliation{JILA and Department of Astrophysical and Planetary Sciences, University of Colorado, Boulder, CO 80309, USA}
\email{jason.dexter@colorado.edu}

\author{Keiichi Asada}
\affiliation{Institute of Astronomy and Astrophysics, Academia Sinica, 11F of Astronomy-Mathematics Building, AS/NTU No. 1, Sec. 4, Roosevelt Rd, Taipei 10617, Taiwan, R.O.C.}

\author[0000-0002-2322-0749]{Christiaan D. Brinkerink}
\affiliation{Department of Astrophysics, Institute for Mathematics, Astrophysics and Particle Physics (IMAPP), Radboud University, PO Box 9010, 6500 GL Nijmegen, The Netherlands}

\author[0000-0002-2526-6724]{Heino Falcke}
\affiliation{Department of Astrophysics, Institute for Mathematics, Astrophysics and Particle Physics (IMAPP), Radboud University, PO Box 9010, 6500 GL Nijmegen, The Netherlands}
\affiliation{
ASTRON, P.O. Box 2, 7990 AA Dwingeloo, The Netherlands
}
\affiliation{
Max-Planck-Institut f\"{u}r Radioastronomie, Auf dem H\"{u}gel 69, D-53121 Bonn, Germany}

\author{Paul Ho}
\affiliation{Institute of Astronomy and Astrophysics, Academia Sinica, 11F of Astronomy-Mathematics Building, AS/NTU No. 1, Sec. 4, Roosevelt Rd, Taipei 10617, Taiwan, R.O.C.}

\author[0000-0001-5037-3989]{Makoto Inoue}
\affiliation{Institute of Astronomy and Astrophysics, Academia Sinica, 11F of Astronomy-Mathematics Building, AS/NTU No. 1, Sec. 4, Roosevelt Rd, Taipei 10617, Taiwan, R.O.C.}

\author[0000-0001-9564-0876]{Sera Markoff}
\affiliation{Anton Pannekoek Institute for Astronomy \& GRAPPA, University of Amsterdam, Postbus 94249, 1090GE Amsterdam, The Netherlands}

\author[0000-0002-2367-1080]{Daniel P. Marrone}
\affiliation{Steward Observatory, University of Arizona, 933 North Cherry Avenue, Tucson, AZ 85721, USA}

\author{Satoki Matsushita}
\affiliation{Institute of Astronomy and Astrophysics, Academia Sinica, 11F of Astronomy-Mathematics Building, AS/NTU No. 1, Sec. 4, Roosevelt Rd, Taipei 10617, Taiwan, R.O.C.}

\author[0000-0002-4661-6332]{Monika Moscibrodzka}
\affiliation{Department of Astrophysics, Institute for Mathematics, Astrophysics and Particle Physics (IMAPP), Radboud University, PO Box 9010, 6500 GL Nijmegen, The Netherlands}

\author[0000-0001-6081-2420]{Masanori Nakamura}
\affiliation{Institute of Astronomy and Astrophysics, Academia Sinica, 11F of Astronomy-Mathematics Building, AS/NTU No. 1, Sec. 4, Roosevelt Rd, Taipei 10617, Taiwan, R.O.C.}

\author[0000-0001-8276-0000]{Alison Peck}
\affiliation{Gemini Observatory, 670 N. A'ohoku Pl., Hilo, HI 96720, USA}

\author[0000-0002-1407-7944]{Ramprasad Rao}
\affiliation{Academia Sinica Institute of Astronomy and Astrophysics, 645 N. A'ohoku Place, Hilo, HI 96720, USA}

\begin{abstract}
We present ALMA observations at 233, 678, and 870 GHz of the 
Galactic Center black hole, Sagittarius A*.  These observations reveal a flat spectrum over this frequency range with spectral index $\alpha \approx -0.3$, where the flux density $S \propto \nu^\alpha$.  We model the submm and far infrared spectrum with a one zone synchrotron model of thermal electrons. We infer electron densities $n = (2-5) \times 10^6$ cm$^{-3}$, electron temperatures $T_e = (1-3) \times 10^{11}$ K, and magnetic field strength $B = 10-50$ G. The parameter range can be further constrained using the observed quiescent X-ray luminosity. The flat submm spectrum results in a high electron temperature and implies that the emitting electrons are efficiently heated. We also find that the emission is most likely optically thin at 233 GHz. These results indicate that millimeter and submillimeter wavelength very long baseline interferometry of \sgra\ including those of the Event Horizon Telescope should see a transparent emission region down to event horizon scales.
\end{abstract}


\section{Introduction}

The Galactic center compact radio source, Sagittarius A* \citep[Sgr A*,][]{balick1974} is
the prototype for low-luminosity accretion onto a massive black
hole \citep{2014ARA&A..52..529Y}. Its inverted radio spectrum rises to a submillimeter (submm) or
far-infrared peak \citep{falcke1998,bower2015}. The radio source varies with
an rms rising from $\simeq 10\%$ in the radio \citep{2004AJ....127.3399H,2006ApJ...641..302M,bower2015}
to $\simeq 30\%$ at 230
GHz \citep{zhao2003,marrone2008,eckart2008,dexter2014} to an order of magnitude in the near-infrared and factors of a
hundred or thousand in X-rays \citep{doddseden2011,2015ApJ...799..199N,witzel2018}.   Millimeter and submillimeter wavelength linear and circular polarization measurements have provided important diagnostics of the emitting plasma and the accretion flow on scales out to the Bondi radius \citep{aitken2000,1999ApJ...523L..29B,2003ApJ...588..331B,2006ApJ...646L.111M,2007ApJ...654L..57M,2012ApJ...745..115M,2018ApJ...868..101B}.  The emission size decreases
with wavelength \citep{krichbaum1998,2005Natur.438...62S,bower2006,2014ApJ...790....1B,johnson2018}, with a
size $\simeq 40-50 \mu$as at 230 GHz corresponding to roughly 8
gravitational radii ($r_g=GM/c^2$)
\citep{krichbaum1998,doeleman2008,2018ApJ...859...60L}, making Sgr A* a prime
target for studying accretion and strong gravity on event horizon
scales \citep{falcke2000shadow,2010ApJ...718..446J}. The first such test was recently performed
with the discovery of near-infrared flares orbiting the black hole at $\simeq 6-10 r_g$
\citep{gravityflare}.  Event Horizon Telescope imaging of the black hole in M87 demonstrates the capability for similar imaging of Sgr A* \citep{2019ApJ...875L...1E,2019ApJ...875L...2E,2019ApJ...875L...3E,2019ApJ...875L...4E,2019ApJ...875L...5E,2019ApJ...875L...6E}.

Intensive studies of Sgr A* from radio to X-ray wavelengths provide tests of accretion
\citep{melia1998,narayan1995,quataert1999,ozel2000,yuan2003} and outflow
\citep{falcke2000jet} models. The development of general
relativistic MHD simulations of black hole accretion flows
\citep[GRMHD,][]{devilliers2003,gammie2003} has led to a large
effort in comparing those models to data, including the variable submm spectral energy distribution 
(SED)
\citep[e.g.,][]{noble2007,dexter2009,dexter2010,moscibrodzka2009,moscibrodzka2013,shcherbakov2012,2013MNRAS.431.2872D,chan2015image}.  

One of the most important constraints for the models is the location and spectral shape
near the peak of the SED. Past observations have
characterized the time variable SED, but with only a few  simultaneous measurements in multiple submm bands \citep{marronephd}. These
measurements along with recent data from ALMA
\citep{bower2015,liu2016492} and detections of variable flux from Sgr A*
in the far-infrared \citep{stone2016,vonfellenberg2018} suggest that
the peak lies somewhere in the THz range. The submm bump is also found to be less peaked than previously thought, which implies a higher
electron temperature and an optically thin emission region near the peak of the SED.  

Here we report flux density measurements from ALMA observations of Sgr
A* simultaneous at 233 and 678 GHz, as well as a precise measurement
at 868 GHz, the first at that frequency using an interferometer
(Section~\ref{sec:observations}). Interferometric observations at THz frequencies
have the advantage of high angular resolution over single dish observations, which is important for 
separating the compact source from the extended and bright Galactic Center
emission.  In cases where phase self-calibration is possible, interferometers also provide better calibration through rejection of temporally and spatially variable atmospheric emission.
We show that the simultaneous 233 and 678 GHz
measurements are consistent 
with and more precise than earlier ones using the SMA. The 868 GHz
flux density is somewhat lower than previously found at 850 GHz with
the CSO, possibly as the result of unsubtracted extended flux density. From our data, we show that the spectral peak occurs at a
frequency $\gtrsim 900$ GHz. Combined with upper limits from Herschel
SPIRE and PACs, we use a one zone model of synchrotron radiation from a
thermal population of electrons to infer the source properties (Section~\ref{sec:discussion}). We
show that the spectral peak occurs at $\simeq 1-2$ THz, and that the
emission region is likely optically thin for frequencies $\gtrsim 230$ GHz. 

\section{Observations, Data Reduction, and Results}
\label{sec:observations}

Observations of \sgra\ were obtained on two days in March 2017 as part of ALMA Cycle 4.  
On 18 March 2017, observations were obtained in Band 10 (868 GHz).
Weather was excellent for the Band 10 measurements with  0.29 mm precipitable water vapor (PWV).
On 22 March 2017, observations were obtained in Bands 6 and 9 (233 and 678 GHz, respectively) within 45 minutes of each other.  
For all three bands, integrations on \sgra\ were 2 minutes in each
band; calibrator integrations were of comparable duration.  

Observations in each band were obtained in standard correlator configurations with
four spectral windows (SPWs) each with 2 GHz bandwidth and 128 channels.  Data were obtained in two 
orthogonal linear polarizations but correlations were computed only for 
parallel hands.  

\begin{figure}
\includegraphics{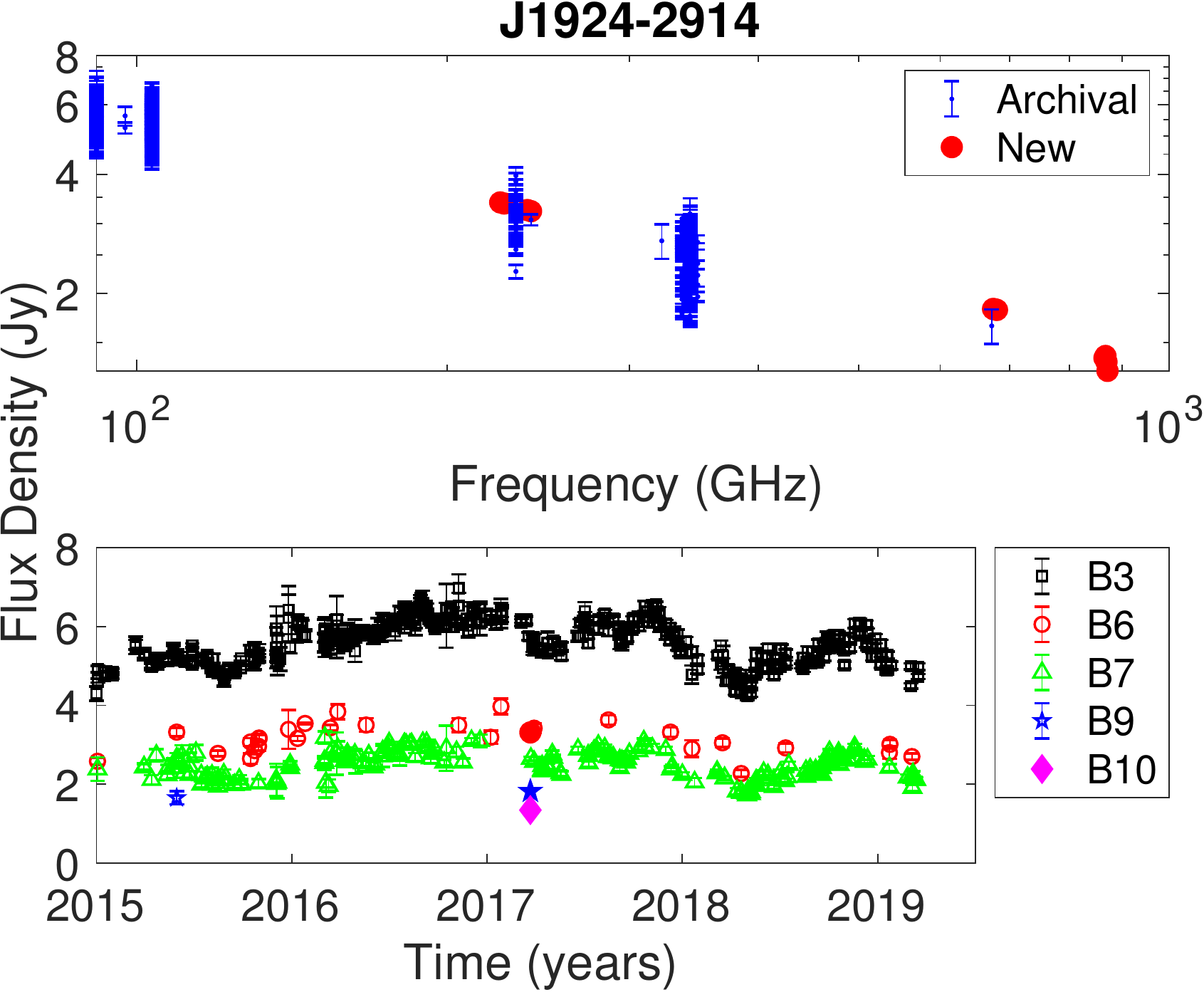}
\caption{Flux density measurements for the flux calibrator
J1924-2914 from these observations and archival ALMA data from 
January 2015 to March 2019.  The top panel shows the spectrum of all measurements.  Filled red
circles are the new measurements and blue dots are the archival data.  The bottom panel shows the 
light curves for Bands 3, 6, 7, 9, and 10 corresponding roughly to 90, 230, 345, 678, and 868 GHz, respectively.  Filled symbols are the new Band 6, 9, and 10 measurements.
\label{fig:J1924}
}
\end{figure}

Data reduction was performed using CASA, following standard procedures for flux, gain, and bandpass
calibration, including phase self-calibration on short time scales for \sgra\ and each calibrator.
Flux calibration is based on estimates of the flux density of the ALMA gain calibrator J1924-2914, which was observed primarily in Bands 3 and 7 (90 GHz and 345 GHz, respectively) and 
then extrapolated to our observing bands.  In \autoref{fig:J1924}, we compare the measured flux
density on J1924-2914 against archival measurements.  
Comparisons to archival ALMA Band 6 measurements for all calibrators show excellent consistency
with differences to nearest measurements $\lsim 10\%$.
There are no Band 9 and 10 measurements within one year of our measurements for any of the
calibrators.  There are a pair of Band 9 measurements for J1924-2914 from two years prior
that agree within 10\% of the extrapolated flux (and resultant measurement).  
The measured Band 10 flux density of J1751+0939 $S=1.31 \pm 0.01$ Jy agrees with the ALMA
calibrator database Bands 3 and 7 extrapolated flux density of $S=1.34$ Jy.
We estimate that systematic flux density errors in Bands 9 and 10 are $\lsim 20\%$.

\begin{deluxetable}{lrrrr}
\tablecaption{Flux Measurements and Spectral Indices \label{tab:flux}}
\tablehead{
\colhead{Source} & \colhead{$S_{233}$} & \colhead{$S_{678}$} & \colhead{$S_{868}$} & \colhead{$\alpha$} \\
                 & \colhead{(mJy)} &\colhead{(mJy)} &\colhead{(mJy)} & \\               
}
\startdata
J1700-2610   & \dots & \dots & $ 0.426 \pm 0.019 $ & \dots \\ 
J1733-1304   & $ 1.592 \pm 0.045 $ & \dots & \dots & \dots \\ 
J1733-3722   & \dots & $ 0.353 \pm 0.009 $ & \dots & \dots \\ 
J1744-3116   & $ 0.270 \pm 0.008 $ & $ 0.096 \pm 0.006 $ & $ 0.058 \pm 0.009 $ & $ -1.11 \pm  0.07 $ \\ 
J1751+0939   & \dots & \dots & $ 1.311 \pm 0.008 $ & \dots \\ 
J1924-2914   & $ 3.312 \pm 0.086 $ & $ 1.819 \pm 0.004 $ & $ 1.342 \pm 0.051 $ & $ -0.65 \pm  0.03 $ \\ 
Sgr A*       & $ 2.886 \pm 0.043 $ & $ 2.183 \pm 0.026 $ & $ 1.864 \pm 0.067 $ & $ -0.31 \pm  0.02 $ \\ 
\enddata
\tablecomments{$\alpha$ is determined over all three frequency bands.  Note that the 233 and 678 GHz observations were obtained on the same day but 868 GHz observations were obtained on a different day.}
\end{deluxetable}

Images of \sgra\ in Bands 9 and 10 were point sources, while  
Sgr A West is apparent in the Band 6 data.  Given that these are all just a few minute snapshots they do
not present very interesting opportunities for imaging.  \replaced{Image RMS in Bands 6, 9, and 10 were 4, 5, and 14 mJy.}{The rms noise levels in the images in Bands 6, 9, and 10 were 4, 5, and 14 mJy, respectively.}
The array was in a compact configuration with maximum baseline of 2.4 km.
This produced a naturally-weighted synthesized beam of $\sim 1.5$ arcsec in Band 6, $\sim 0.5$ arcsec in Band 9, and
$\sim 0.4$ arcsec in Band 10.

\begin{figure}
  \includegraphics[width=\textwidth]{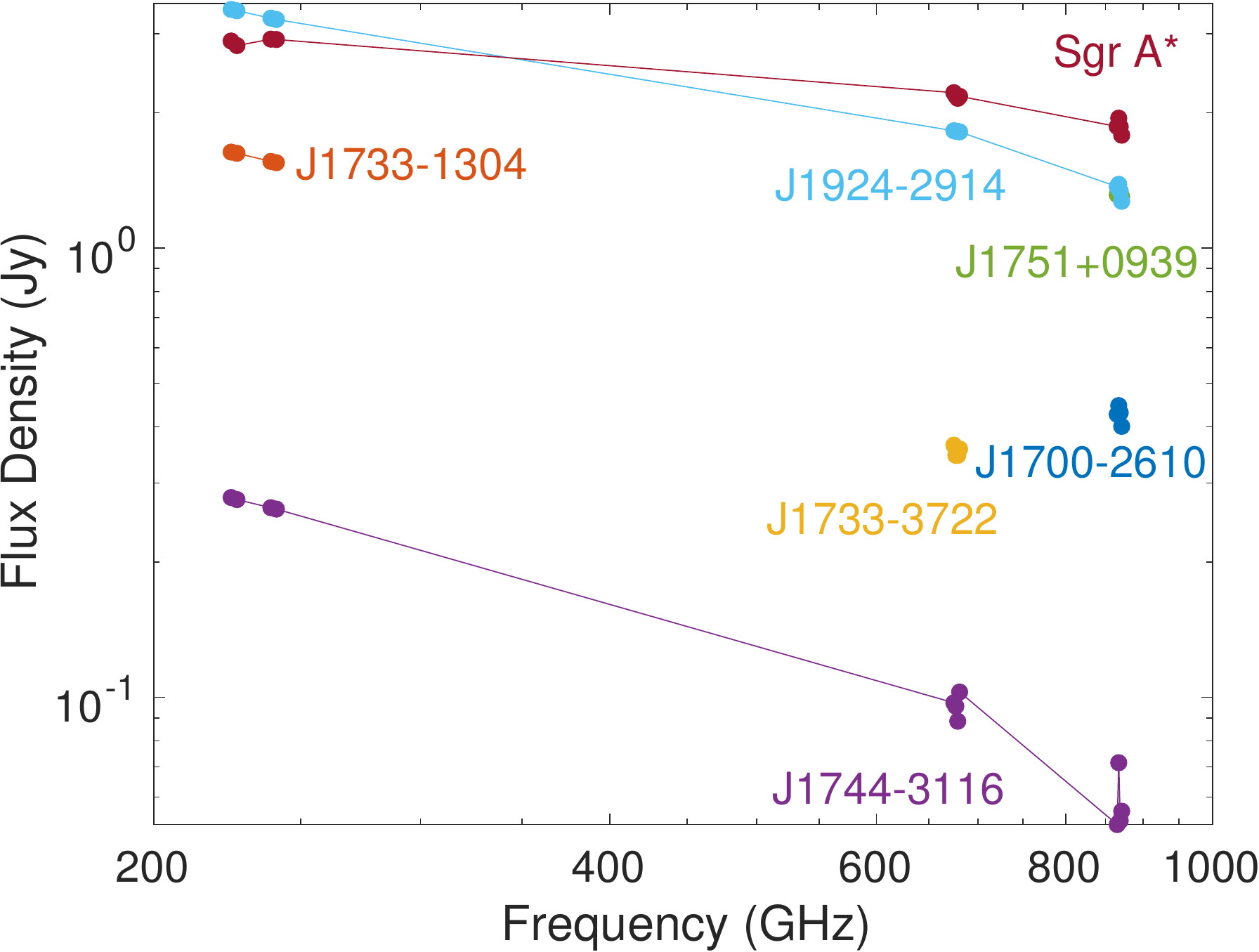}
\caption{Flux densities for \sgra\ and calibrators from ALMA observations on 18 and 22 March 2017.  Flux densities at 870 GHz for J1924-2914 and J1751+0939 are nearly identical and so overlap
in the plot.
\label{fig:allsources}}
\end{figure}

Flux densities were fit for each source in each spectral window using a point source model in the 
visibility domain. \autoref{fig:allsources} shows all flux densities measured.  In \autoref{tab:flux},
we report mean flux densities in each band.  Errors are computed from the scatter in measurements, which 
provides more accurate assessment of errors than propagation of statistical uncertainties.  

We also
compute the spectral index $\alpha$ (using $S \propto \nu^{\alpha}$) for sources with measurements in
all three bands.
The spectrum of \sgra\ is close to flat with a spectral index
$\alpha=-0.3$ across all 3 bands.  
In comparison, the assumed spectrum of J1924-2914 and the measured
spectrum of J1744-3116 are both steeper with $\alpha=-0.7$ and
$\alpha=-1.1$, respectively. Considering only the simultaneous 233 and 678 GHz \sgra\ data, $\alpha = -0.26 \pm 0.02$.

\begin{figure}
\includegraphics[width=0.8\textwidth]{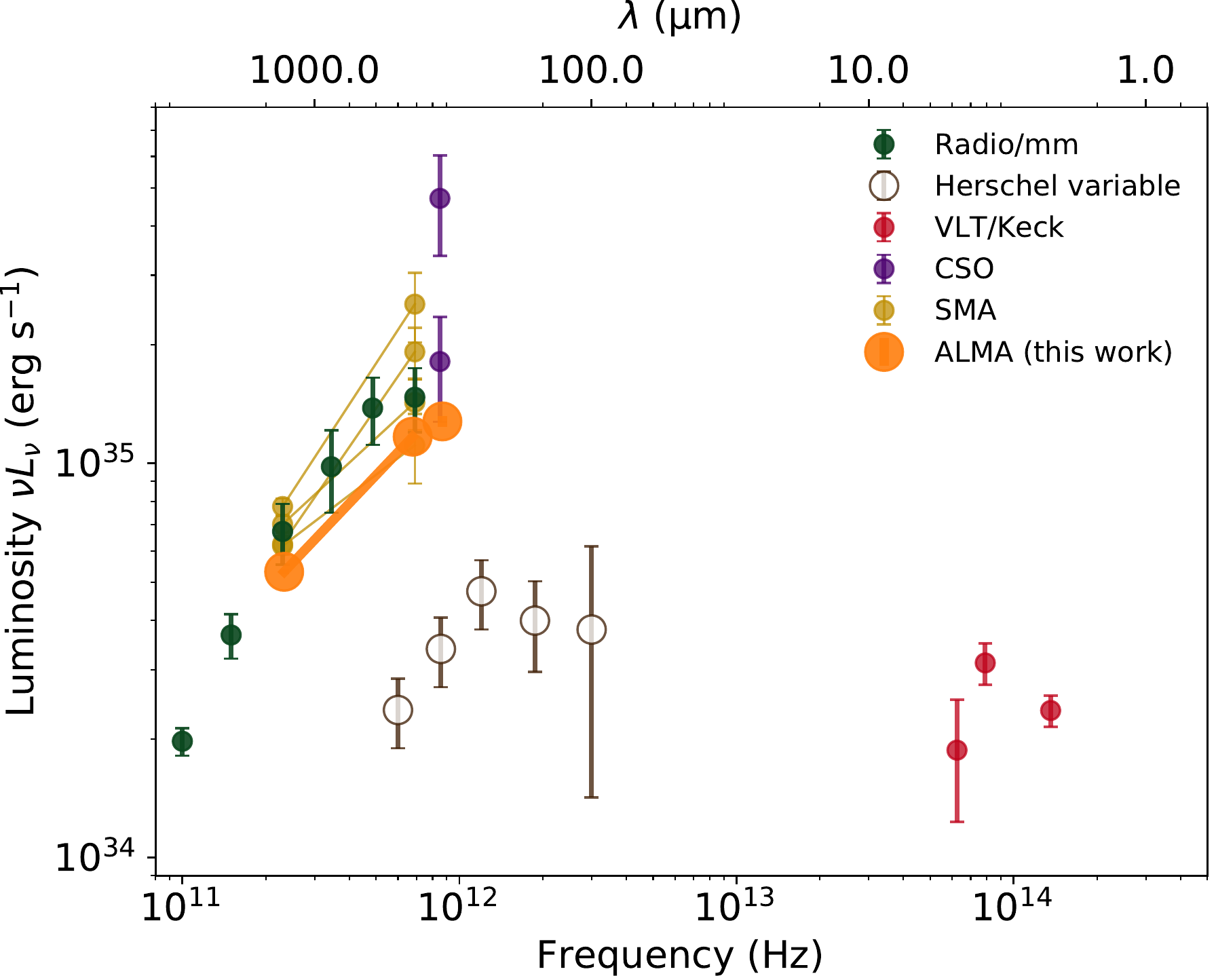}
\caption{The mm to NIR spectrum of \sgra. Orange circles show the new
  ALMA measurements, with errors much smaller than the symbol
  size. The green circles show selected radio to submm data
  \citep{falcke1998,brinkerink2015,bower2015,liu2016492,liu2016pol}. The yellow
  circles connected with thin lines are simultaneous SMA data
  \citep{marronephd}. Single dish measurements from CSO at 850 GHz
  \citep{serabyn1997,yusefzadeh2006} are shown in purple. Detections of variable flux from Sgr A* with Herschel SPIRE \citep[250, 350, 500  micron,][]{stone2016} and
  PACS \citep[160 and tentatively 100 micron,][]{vonfellenberg2018} are shown as  brown open symbols under the assumption that they represent lower limits on the flux density at these frequencies. Near-infrared median flux
  density measurements \citep{schoedel2011,doddseden2011,witzel2018}
  are shown in red.
\label{fig:spectrum}
}
\end{figure}

\begin{figure}
\includegraphics[width=0.8\textwidth]{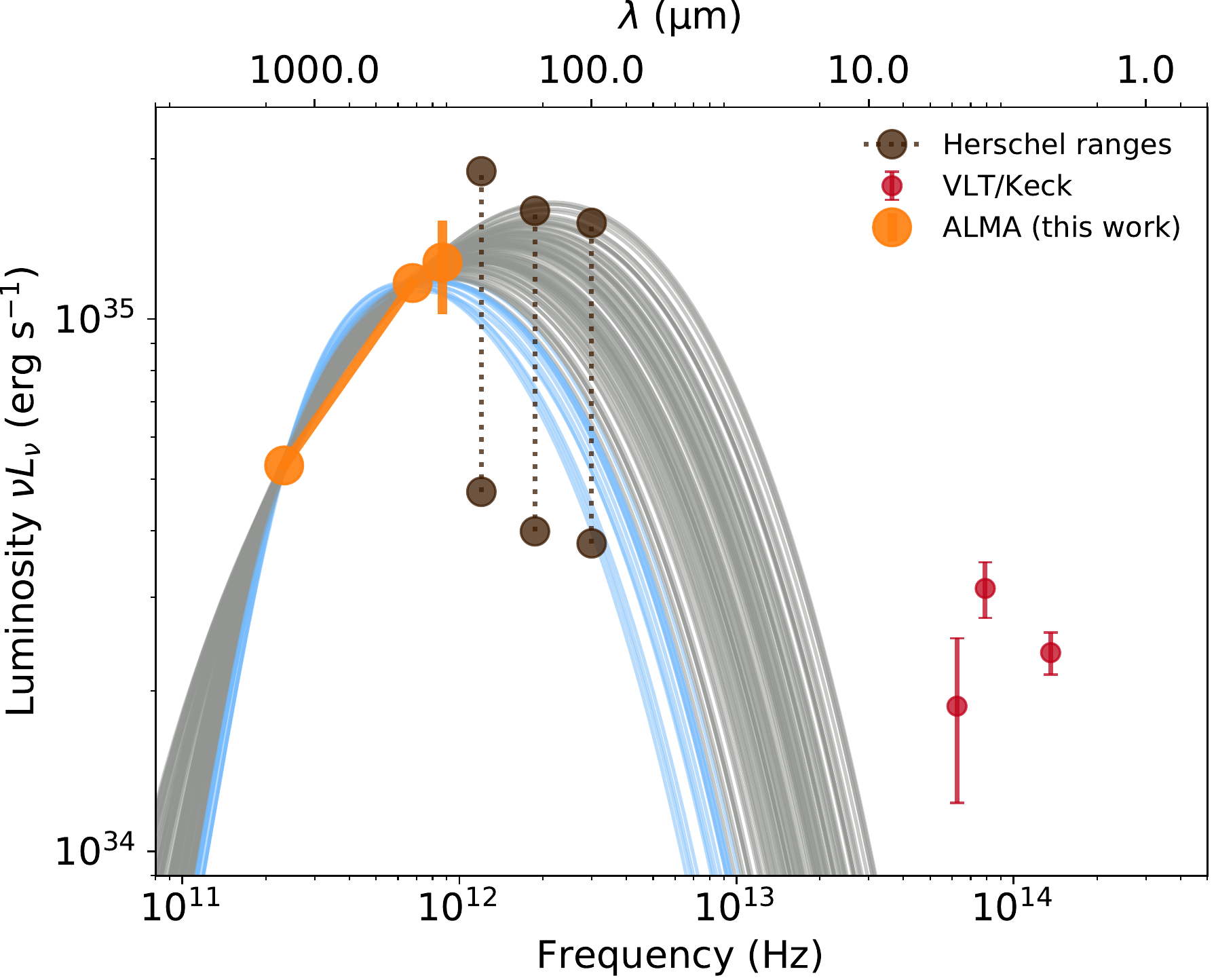}
\caption{Sample one zone model fits (lines) to our new ALMA
  measurements and derived flux density ranges from recent Herschel detections of
  variable flux. The SED peak occurs at 1-2
  THz with a bolometric luminosity $\simeq 5 \times
  10^{35}\,\ergs$. The emission region is usually optically thin (gray) rather than optically thick (blue) at 230 GHz. The 68\% confidence interval ranges for the plasma
  parameters are: $n = (2-5) \times 10^6 \hspace{2pt} \rm cm^{-3}$, $T_e =
  (1-3) \times 10^{11}$ K, $B = 10-50$ G. We note that these parameters
  are strongly correlated and depend on the chosen emission region size.
\label{fig:onezone}
}
\end{figure}

\begin{figure}
\includegraphics[width=0.8\textwidth]{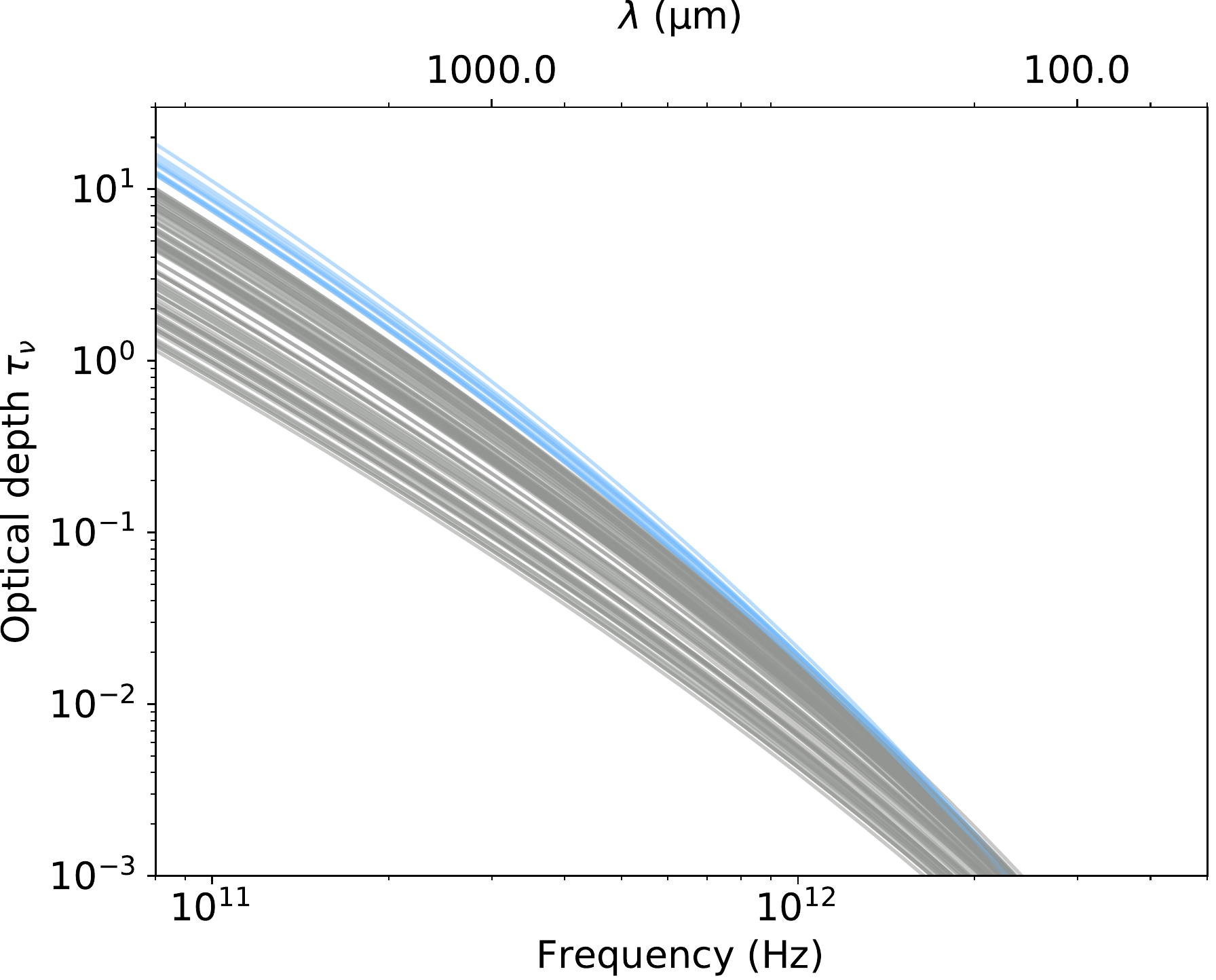}
\caption{Optical depth $\tau_\nu$ as a function of frequency for a sample of our one zone model fits. All models are optically thin at the submm/THz spectral peak, while a small fraction (shown in blue) can be marginally optically thick ($\tau_\nu > 1$) at 230 GHz.
\label{fig:taunu}
}
\end{figure}

\section{Discussion and Conclusions}
\label{sec:discussion}

We plot our new ALMA submm spectrum of \sgra\ along with
measurements from the radio to the NIR
(\autoref{fig:spectrum}). The measured flux density at 230 and 678 GHz are at the low end of the range characterized in previous work
\citep[e.g.,][]{marronephd,dexter2014,bower2015,liu2016pol}. Previous
THz single dish measurements with CSO found higher flux
densities \citep{serabyn1997,yusefzadeh2006}.  The difference with these 
earlier measurements may be the result of variability or calibration uncertainties 
associated with THz single dish measurements in this environment with significant extended emission.  The characteristic time scale of variability for \sgra\ at 230 GHz and higher frequencies has been measured to be $\tau \approx 8_{-4}^{+3}$ hr \citep{dexter2014}.  Thus, 
the 45-minute separation between 230 and 678 GHz measurements is nearly simultaneous while the four-day separation with the 868 GHz measurements is significantly longer than the variability coherence time.  Long time scale rms variability is approximately 20\% \citep{2015ApJ...802...69B}, which is comparable to the systematic error in the 680-GHz flux density that we estimate.
Accordingly, the spectral index of the simultaneous 233 and 678 GHz measurements is the strongest spectral constraint. Still, we note the overall consistency of a power-law spectral index between 233 and 868 GHz in these data.

Our results for \sgra\ are among the best characterized spectrum of any low luminosity AGN (LLAGN) and show one of the flattest spectra for these sources. \citet{2011AJ....142..167D} characterize the centimeter-to-millimeter wavelength spectra of 21 LLAGN, including 5 with simultaneous data at 100 and 150 GHz, finding flat or inverted spectra for many sources but with no contemporaneous data at frequencies above $>150$ GHz.
ALMA observations of M87 extend to 650 GHz and indicate a steep spectrum at frequencies above $\sim 200$ GHz \citep{prieto2016}.
In the case of M94, \citet{2017MNRAS.468..435V} found a flat spectrum up to 100 GHz but place no strong constraints on the spectrum between 100 GHz and the optical as the result of stellar confusion.  Contemporaneous observations of the black hole in M81 also indicate a flat spectrum up to 350 GHz but the detailed spectrum is difficult to characterize due to the absence of high angular resolution submillimeter and infrared observations as well as short time scale variability \citep{2008ApJ...681..905M,bower2015}.  \citet{2008A&A...483..741I} find for the nuclear region of Cen A, a spectral index of $\alpha = -0.2$ to -0.6 between 90 and 230 GHz.  Similarly, \citet{2017ApJ...843..136E} find a flat spectrum for Cen A
between 350 and 698 GHz with non-simultaneous ALMA observations.  ALMA THz spectra of a wider sample of LLAGN are necessary to characterize this population and assess their viability for high frequency imaging.

\added{The spectrum of Sgr A* must have a steep spectral cutoff between the submm (flux density of Jy) and NIR (flux density of mJy)}. We find that this must occur at frequencies above 900 GHz,
consistent with the previous single dish detections of Sgr A* at 900
GHz. The relatively flat submm SED found by ALMA
\citep{bower2015,liu2016492,liu2016pol} and Herschel SPIRE/PACS \citep{stone2016,vonfellenberg2018} measurements has implications for
the physical properties of the emitting plasma on event horizon
scales. 
We exclude longer wavelength radio and NIR flux densities. The radio emission originates at large radius, where the density, magnetic field strength, and temperature are lower. Both the radio and NIR flux densities may have significant contributions from additional, possibly non-thermal electron populations, which are not included in our one-zone model.  Following \citet{vonfellenberg2018} we estimate the physical
properties of the emission region by fitting a one zone synchrotron
emission model to the new ALMA data as well as implied upper and lower
limits from Herschel. The simultaneous 233 and 678 GHz measurements are
used with their statistical error bars. We adopt an uncertainty of
$20\%$ on the  868 GHz value to account for the (unknown) variability
at that frequency. 

The Herschel detections are of flux variations from Sgr A* on top of a bright background, which are plotted as open circles in \autoref{fig:spectrum}. We further follow \citet{stone2016} and take the detected variable flux densities as lower limits to the median value. That implicitly assumes that the rms variability amplitude is $< 100\%$ (e.g., does not consist of large amplitude flares as observed in the NIR/X-ray). Following \citet{vonfellenberg2018}, we further derive upper limits on the median flux density by assuming a minimum variability amplitude of $25\%$ during the observations (25.5h for SPIRE, 40h for PACS). As a result of these assumptions, the final allowed range in flux density is a factor of 4 at each frequency. We note that  at 350 GHz, the median flux density estimated from SPIRE observations would be $\simeq 2$ Jy, which underestimates the measured value [$3.6 \pm 0.8$ Jy,][]{bower2015}. We expect higher rms variability at higher frequencies where we use these upper limits. Still these measurements are derived quantities and so are less secure than the ALMA data. As discussed below, we find similar (but slightly worse) constraints when leaving out the Herschel data.

We parameterize the emission region as a sphere of constant particle
density $n$, electron temperature $T_e$, and magnetic field strength
$B$. The sphere's angular diameter is set equal to $40 \mu$as \citep{doeleman2008,2018ApJ...859...60L,johnson2018}. We use
a black hole mass of $M = 4.1\times10^6 M_{\odot}$ and a distance of
$D = 8.2$ kpc \citep{2018A&A...615L..15G}. We calculate the observed flux density
accounting for synchrotron emission and absorption from a thermal
population of electrons using the fitting function expressions from Appendix A of \citet{dexter2016}. We neglect all relativistic effects in the spatial
and velocity distribution of the material and on the photon
trajectories. Most critical is Doppler beaming
\citep[e.g.,][]{sunyaev1973}, which broadens the spectrum. We also neglect radiative cooling, which should be negligible for the plasma conditions in Sgr A* \citep{dibi2012}. We sample
the model parameter space using the \texttt{emcee} Markov Chain Monte Carlo code
\citep{emcee}. We use the default sampler with logarithmic priors $n$, $T_e$, and plasma
$\beta = p_g/p_B \propto n/B^2$ where $p_g$ and $p_B$ are the gas and magnetic pressures and we have assumed a constant ion temperature
proportional to the virial temperature. Parameter bounds are $T_e \ge T_b$ the brightness temperature, necessary for obtaining a one zone solution, and $10^{-3} < \beta < 10^{3}$.

Sample model fits are shown in
\autoref{fig:onezone} along with the ALMA data and assumed Herschel ranges used for fitting. The peak of the SED in $\nu L_\nu$ is well
constrained to be at $\nu_p = (1-2) \times 10^{12}$ Hz (all ranges
68\% confidence intervals), close to our new 868 GHz ALMA
measurement. The bolometric luminosity of the 
submm bump is found to be $L_{\rm bol} = (4-6) \times
10^{35}\,\ergs$. This is about a factor of 2 smaller than found in
past work \citep[e.g.,][]{yuan2003}, in part based on the higher flux
densities at 850 GHz from CSO data \citep[see
also][]{vonfellenberg2018}. We find similar results with somewhat larger ranges when leaving out the Herschel data: $\nu_p = (1-3) \times 10^{12}$ Hz and $L_{\rm bol} = (4-8) \times 10^{35}\,\ergs$.

The well constrained $\nu_p$ and $L_{\rm bol}$ lead in turn to
estimates for the plasma properties. We measure these
from the one zone model to be $n = 2-5 \times 10^6 \hspace{2pt} \rm cm^{-3}$, $T_e =
  1-3 \times 10^{11}$ K, $B = 10-50$ G. The associated plasma $\beta
  \simeq 1-100$. Near the peak, all models are optically thin \autoref{fig:taunu}. \replaced{The
  location of the SED peak is set by the exponential cutoff in $j_\nu$ for
  $\nu/\nu_c \gg 1$}{The location of the SED peak is set by the exponential cutoff in $j_\nu \sim e^{-1.9 (\nu/\nu_c)^{1/3}}$ for $\nu/\nu_c \gg 1$} rather than by the transition to an optically thin
  emission region. \added{Our viable models have a range of $\nu/\nu_c = 10-20$ ($\nu_c = 0.1-0.2$ THz).} This results in a broader, flatter spectrum near
  the peak favored by the flat or slowly declining flux density from $233$ to $678$ to $868$ GHz as measured by
  ALMA and past SMA data.

  The derived parameter ranges, particularly for $n$ and $B$, are
  strongly correlated. The critical frequency scales as $\nu_c \propto
  B T_e^2$ and sets the spectral peak, while in the one zone model at
  fixed radius the bolometric luminosity is proportional to the
  synchrotron emissivity near the peak which scales as $j_\nu \sim n
  B^2 T_e^{5/2}$. For the model to produce the observed flux, $T_e > T_b
  \sim 6 \times 10^{10}$ K where $T_b$ is the observed 230 GHz brightness temperature. We see clear correlations as anticipated from the
  forms of $\nu_c$ and $j_\nu$. In particular, the magnetic field strength is
  anti-correlated with both the particle density and electron
  temperature. The spectral shape and our assumed parameter bounds (particularly $\beta < 10^3$) provide some additional
  information, leading to our inferred parameter ranges. Using
  simultaneous 233 and 868 GHz data leads to better constrained parameter
  ranges than the same exercise done in \citet{vonfellenberg2018}. The
  basic results are otherwise identical.

We can break this degeneracy by including an approximate calculation of the 2-10 keV X-ray luminosity from the synchrotron self-Compton (SSC) process \citep[e.g.,][]{falcke2000jet}. We use the method described in \citet{chiaberge1999} and \citet{drappeau2013} to estimate the SSC spectrum. Imposing an upper limit $L_X < 2\times10^{33} \,\rm erg \,\rm s^{-1}$ \citep{baganoff2003} removes all of the higher $\beta$ (high $n$, $T_e$) models where the SSC peak is near the X-ray and the scattering optical depth is highest. This constraint removes about half of the models. The choice of X-ray luminosity upper limit is conservative since the quiescent emission is dominated by the large scale accretion flow. Estimates from the X-ray spatial surface brightness distribution \citep{shcherbakov2010}, variability \citep{neilsen2013}, and spectrum \citep{wang2013} all favor a near horizon component that is a factor of $\gtrsim 10$ smaller. The resulting parameter ranges when including this constraint are $n = 2-3 \times 10^6 \hspace{2pt} \rm cm^{-3}$, $T_e =
  1-2 \times 10^{11}$ K, $B = 20-50$ G. The main improvement is a narrowed range of allowed plasma $\beta \simeq 1-10$.

The resulting electron temperature is higher than in some past RIAF
models where optical depth set the shape of the submm peak
\citep[e.g.,][]{ozel2000,yuan2003,noble2007,huang2009,chan2009,moscibrodzka2009,dexter2010}. The
electron temperature is decoupled from that of the ions since at the
low inferred densities the plasma is collisionless
\citep[e.g.,][]{rees1982}. The ion
temperature near the event horizon is likely close to virial, $k T_i
\simeq 10^{12} (r_g / R)$ K. Here our assumed size is roughly $4 r_g$,
meaning that the implied electron temperature is within a factor of
2-3 of the ion temperature. The emitting electrons are therefore
heated efficiently. This is most easily explained if the magnetic
field is strong (plasma $\beta \gtrsim 1$) in the emission region \citep[e.g.,][]{quataert2000,howes2010,ressler2015,rowan2017,werner2018,kawazura2018}.

The particle density we find is comparable to past estimates from spectral modeling \citep{ozel2000,yuan2003,chan2009,moscibrodzka2009,dexter2010,shcherbakov2012}. It also follows the $n \propto r^{-1}$ scaling seen in Sgr A* from scales of the Bondi
radius down to the event horizon
\citep{baganoff2003,marrone2007,gillessen2019}. For our temperatures and density, the Faraday rotation optical depth \emph{internal} to the emission region is: 

\begin{equation}
\tau_{\rho V} \simeq \frac{2 n e^3 B R} {m_e^2 c^2 \nu^2} \frac{K_0(\theta_e^{-1})}{K_2(\theta_e^{-1})}
\end{equation}

\noindent where $K_n(x)$ is a modified Bessel function and we have used the high-frequency limit $\nu/\nu_c \gg 1$ \citep{jones1979,quataert2000,shcherbakov2008,dexter2016}. For $\tau_{\rho_V} \gg 1$, the linear polarization goes through many oscillations. Small differences $\Delta \tau_{\rho_V} > 1$ across the image will then lead to depolarization \citep[e.g.,][]{agol2000}. At 233 GHz, we find $\tau_{\rho_V} \simeq 0.2-3$. Most of the models should therefore not be depolarized and should be capable of producing the  observed linear polarization of Sgr A* \citep{aitken2000,2003ApJ...588..331B,marrone2008,2018ApJ...868..101B}. For these parameters, the Faraday conversion effect is about an order of magnitude weaker.

The emitted fractional linear and circular polarization are $\simeq 60\%$ and $\simeq 3.5\%$ for our fiducial parameters and an angle between the line of sight and magnetic field of $\theta = \pi/6$. The source must be somewhat beam \citep[e.g.,][]{bromley2001} or Faraday \citep[e.g.,][]{shcherbakov2012,dexter2016} depolarized. The observed $\simeq 1\%$ mm-wavelength circular polarization \citep{2012ApJ...745..115M,2018ApJ...868..101B} could arise from either direct emission or Faraday conversion.

For simplicity here we have considered one zone models. State of the art radiative models based on GRMHD simulations in general produce ranges of densities, field strengths, and electron temperatures near the black hole. When the electrons are efficiently heated everywhere ($T_i/T_e \simeq$ constant) the emission is dominated by the densest material near the midplane of the accretion flow \citep[e.g.,][]{moscibrodzka2009,dexter2010,shcherbakov2012,drappeau2013}. In that case, the physical conditions are similar to those of the one zone model. In low magnetic flux (SANE) models where electron heating strongly depends on the plasma $\beta$, the model is effectively composed of two zones: a dense accretion flow with cold electrons that do not radiate much in the submm, and a more tenuous jet boundary (or funnel wall) with hot electrons that produce the observed emission \citep{moscibrodzka2014,chan2015image,ressler2017}. In that case, our inferred physical conditions apply to the jet wall region producing the observed radiation. In particular, the submm emission may be depolarized in the two zone model from passing through the dense, cold accretion flow \citep{moscibrodzka2017,jimenezrosales2018}. It remains to be seen whether such models can match the high submm linear polarization fraction seen in Sgr A*.

We have further assumed a thermal electron distribution function. Using a power law shape yields similar parameter estimates and spectral shape, with steep slopes $p \gtrsim 4$ (high frequency spectral index $\alpha \gtrsim 3/2$), minimum electron energies $\gamma_{\rm min} \sim 100$, and magnetic field strengths $B=10-50$ G. In particular, we have not found one zone thermal models which can fit both the submm spectral peak and the median flux density in the near-infrared.

The broad spectral shape peaking in the THz regime imply a mostly optically thin emission region at 233 GHz. Approximately $90\%$ of the
sampled models have $\tau_\nu < 1$ and all have $\tau_\nu < 2$
at that frequency (\autoref{fig:onezone}). All models are optically thin at 345 GHz. Theoretical models like those discussed above 
generally find that the optical depth varies substantially across the
observed image due to varying fluid properties and to Doppler beaming effects\citep[e.g.,][]{broderick2006}. Still, our findings suggest that mm-VLBI observations with the EHT should be able to see a
mostly transparent emission region down to event horizon scales. The absence of a steep spectral cutoff establishes the possibility of higher frequency VLBI observations, either from the ground or space, that would achieve extraordinary angular resolution \citep{2018cosp...42E1035F,2019arXiv190309539F}.

\acknowledgements
JD thanks S. von Fellenberg for providing the one zone code used to infer source parameters and to make \autoref{fig:onezone}, as well as for helpful comments which improved the manuscript. This paper makes use of the following ALMA data: ADS/JAO.ALMA\#2016.1.00901.S. ALMA is a partnership of ESO (representing its member states), NSF (USA) and NINS (Japan), together with NRC (Canada), MOST and ASIAA (Taiwan), and KASI (Republic of Korea), in cooperation with the Republic of Chile. The Joint ALMA Observatory is operated by ESO, AUI/NRAO and NAOJ.
The National Radio Astronomy Observatory is a facility of the National Science Foundation operated under cooperative agreement by Associated Universities, Inc. JD was supported by a Sofja Kovalevskaja award from the Alexander von Humboldt foundation. SM is supported by an NWO VICI grant (no. 639.043.513).

\facilities{ALMA}
\software{CASA, MATLAB}


\end{document}